\documentclass[10pt,twocolumn,reqno, aps,prb,superscriptaddress]{revtex4-1}%
\usepackage{amsmath}
\usepackage[pdftex]{graphicx}
\usepackage{float}
\usepackage{color}
\usepackage{rotating}
\usepackage[breaklinks=true,colorlinks,citecolor=blue,linkcolor=blue,urlcolor=blue]%
{hyperref}
\usepackage{graphicx}
\usepackage{natbib}
\usepackage{dcolumn}
\usepackage{bm}
\usepackage{hyperref}
\usepackage[caption=false]{subfig}
\usepackage{epstopdf}
\usepackage{ulem}
\usepackage{amsfonts}
\usepackage{amssymb}%
\providecommand{\U}[1]{\protect\rule{.1in}{.1in}}
\begin{document}

\title{Topologically nontrivial magnonic solitons}

\author{Mehrdad Elyasi}
\affiliation{Institute for Materials Research, Tohoku University, 980-8577 Sendai, Japan    
}
\author{Koji Sato} 
\affiliation{Institute for Materials Research, Tohoku University, 980-8577 Sendai, Japan
}
\author{Gerrit E. W. Bauer}%
\affiliation{Zernike Institute for Advanced Materials, University of Groningen, The Netherlands}
\affiliation{Institute for Materials Research $\text{\&}$ AIMR \& CSRN, Tohoku University, 980-8577 Sendai, Japan}

\begin{abstract}
{The intrinsic non-linearities of the spin dynamics in
condensed matter systems give rise to a rich phenomenology that can be strongly
affected by topology. Here we study formation of magnonic solitons in the
topologically nontrivial bandgap of a spin lattice realization of the Haldane
model, in both static and dynamic (Floquet) regimes. We consider
nonlinearities caused by magnetic crystalline anisotropy and magnon-magnon
interactions. We find soliton formation power thresholds as a function of
anisotropy coefficient and interaction strength. We predict different classes
of topological solitons for the same topological class of the underlying
lattice and explain it in terms of a transition from a topologically
nontrivial mass to a trivial one. Our findings imply that a soliton can
phase-separate, containing boundaries between topologically trivial and
non-trivial phases, which is associated with a vanishing spin wave gap.}

\end{abstract}
\maketitle


\section{\label{Intro}Introduction}

In physics, topological equivalence classes refer to Hamiltonians with energy
gaps that with smooth changes in parameter space do not become gapless unless
a quantum phase transition occurs. The robustness with respect to
perturbations \cite{Hasan2010,Qi2011}, as well as the potential for quantum
computation \cite{Nayak2008} made topology an important subfield of modern
condensed matter physics. Topology manifests itself, for instance, in the
quantum Hall effects
\cite{Klitzing1980,Laughlin1983,Haldane1988,Kane2005,Bernevig2006,Bernevig2008,Regnault2011}%
, 3D topological band insulators \cite{Fu2007,Zhang2009,Hasan2010,Qi2011}, and
quantum spin liquids \cite{Lee2006,Balents2010,Yan2011,Cincio2013}. Chiral, or
in the case of the quantum spin Hall effect, helical \cite{Kane2005} edge
modes are caused by bands with nontrivial global phases derived from the
topology of the bulk material. Topology affects the wave functions of
electrons \cite{Konig2007,Regnault2011,Qi2011}, photons
\cite{Haldane2008_1,*Khanikaev2013}, phonons \cite{Yang2015}, magnons
\cite{Shindou2013,*Shindou2013_1}, and quasi-particles such as Majorana
fermions \cite{Fu2009}.

In continuous media, the Damon-Eshbach (DE) surface modes in ferromagnetic
films are chiral \cite{Damon1961,Stancil2009}, but not proven yet to be
topological. Nevertheless, the topology (i.e. the Berry curvature) of magnon
bands in perpendicular magnetized films has been revealed in phenomena such as
the magnon Hall effect and the rotation of wave packets at the edges
\cite{Onose2010,Matsumoto2011}. Motivated by the established topological band
theory \cite{Qi2011}, topological chiral magnonic edge modes in gapped
magnonic crystals have attracted attention. Spin wave dispersions in
ferromagnets are governed by both dipolar and exchange interactions. In a
lattice with nonuniform equilibrium magnetization, the former can break the
inversion and time reversal symmetry, leading to bands with nonzero Chern
number. This implies emergence of chiral edge modes at the boundary of the
lattice with vacuum (or a lattice with different topology)
\cite{Shindou2013,*Shindou2013_1}. The exchange interaction can also be
utilized to design magnonic analogues of static or Floquet-type
\cite{Shirley1965,Scholz2010,Kitagawa2010} Haldane \cite{Haldane1988} spin
lattice models \cite{Zhang2013,Mook2015,Kim2016,Owerre2016,Owerre2017}. The
periodic Floquet variable can be the time or a spatial coordinate. In 2D
spatial lattices, e.g., time \cite{Kitagawa2010} as well as the third spatial
coordinate normal to the 2D plane \cite{Rechtsman2013} have been employed as
the Floquet dimension.

In the case of Fermions, the chiral (or helical) edge modes can be populated
at all temperatures when the chemical potential falls into a non-trivial band
gap. Magnons and other Bosonic edge states, on the other hand, tend to be
empty at low temperatures, which makes experimental observation challenging.
This motivated proposals to utilize non-linear interactions that can drive
edge mode instabilities \cite{Galilo2015} or lead to self-localized wave
packets with a sense of chirality inherited from the original edge modes
\cite{Lumer2013,Ablowitz2014,Leykam2016}.

Solitons are shape-preserving, self-localized modes in dispersive media. In
magnets, solitonic textures such as domain walls, vortices and skyrmions that
exist in equilibrium can be topologically protected \cite{Kosevich1990}. They
also emerge as robust excited states, in which nonlinear interactions
compensate the wave packet dispersion. The latter type of solitons have been
explored theoretically and experimentally in both continuous and discrete
systems such as Bose-Einstein condensates of cold atoms in optical potentials
\cite{Zobay1999,Trombettoni2001,Louis2003,Eiermann2004} as well as light in
fibres and photonic crystals
\cite{Drummond1993,Mitchell1997,Stegeman1999,Wang2007,Lumer2013,Leykam2016}.
Solitons in thin magnetic films can be excited by microwaves
\cite{Slavin1994,Buttner2000,Wu2006,Stancil2009}. {Subsequently, magnonic solitons have been generated in
nano-contacts by spin-transfer \cite{Kaka2005,Slavin2009} and spin-orbit
torques \cite{Demidov2012}. When the spin current is injected locally into an
in-plane magnetized film above a certain threshold, the exchange dispersion is
compensated by a focusing nonlinearity, and a self-localized non-propagating
spin-wave mode emerges at a frequency below the predictions of linear theory
\cite{Slavin2005}, which is referred to as \textquotedblleft spin wave
bullet\textquotedblright}. Droplets of magnon condensates as
generated by parametric pumping can be interpreted as solitons as well
\cite{Demidov2008,Rezende2010,Malomed2010}.

Here we show that the non-linearities generated by crystalline magnetic
anisotropy and magnon-magnon interactions can generate magnonic solitons that
show signatures of the topology of the underlying spin wave band structure of
magnonic crystals with nontrivial band gaps. We search for solitons in the
bulk of static and time-periodic (Floquet) magnonic equivalents of the Haldane
model, i.e. a hexagonal lattice with $C_{6v}$ point group symmetry, but broken
time-reversal symmetry \cite{Haldane1988}. We chose the Haldane model as a
minimal but generic model with a single band gap but nontrivial topology. The
results can be extended to other lattices with band gaps of similar topology.
We attribute the different solitons phases in parameter space to distinct
topologies, even mixed topologies. For the latter case we predict the
existence of interfaces between topologically trivial and non-trivial highly
excited phases.

The paper is organized as follows. Section \ref{sec1} addresses solitons in
the topologically nontrivial band gap of a static magnonic Haldane model.
Section \ref{sec1_a} describes the model and the numerical method used to find
solitons. In section \ref{sec1_b} we show calculated soliton phase diagrams,
and explain them in Sec. \ref{secq_c}. In section \ref{sec2} we focus on the
soliton phase diagram in the Floquet equivalent of the Haldane model. Finally,
in Sec. \ref{sec3}, we evaluate the experimental feasibility and propose two
methods to test the findings of this paper.

\section{\label{sec1}Solitons in the static magnonic Haldane model}

\subsection{\label{sec1_a} Model}

\begin{figure}[t]
\includegraphics[width=0.5\textwidth]{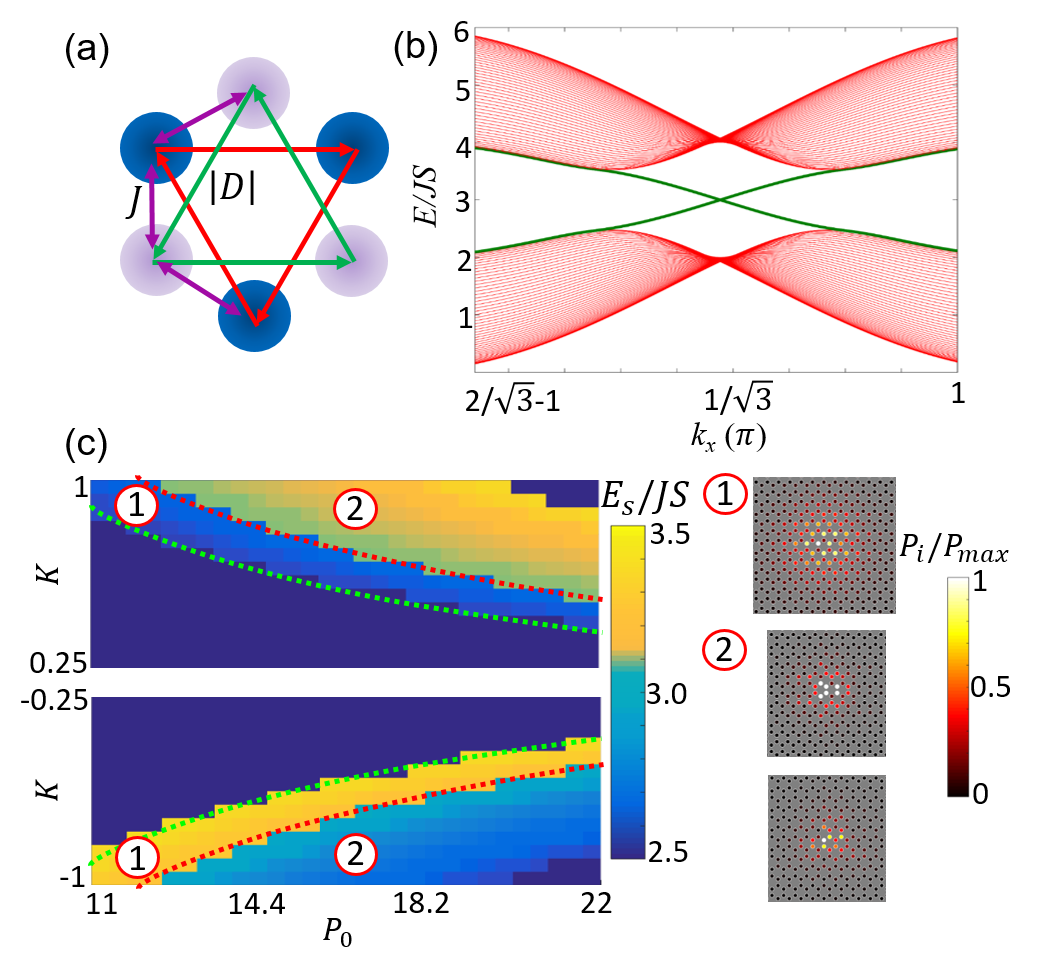}\caption{(a) Schematics of
the lattice, the Heisenberg exchange interactions (purple lines), and the NNN
DMI interactions (green ($v_{ij}=1$) and red ($v_{ij}=-1$) lines). (b) The
bandstructure of a 80 sites wide, quasi-1D ribbon with $J$ = 0.1, $D$ = 0.01,
$K=0$, and $S$ = 10. The edge states (green lines in the gap) merge with the
band edges for large wave numbers. (c) Soliton phase diagram in the space of
anisotropy constant $K$ and integrated intensity $P_{0}$. In the dark-blue
area no solution was found. The color map indicates the energy $E_{s}$ of the
calculated solitons. 1 and 2 label the distinct parts of the phase diagram.
The right panels display representative soliton density distributions found in
1 (the top one) and 2 (the two bottom ones, which are basically the same for
both signs of $K$). Green and red dashed lines mark the thresholds $P_{c,1}$
and $P_{c,2}$, respectively, as detailed in the text. In the color code of the
right panels, $P_{max}\lesssim1.5\,S$ is the peak value of $P_{i}=\langle
\Psi_{s}|n_{i}|\Psi_{s}\rangle$. }%
\label{fig1}%
\end{figure}

The equivalent of the Haldane model \cite{Haldane1988} for magnons can be
derived from the Hamiltonian \cite{Owerre2016,Kim2016},
\begin{equation}
H_{S}=\sum_{\langle i,j\rangle}J\vec{S}_{i}\cdot\vec{S}_{j}+\sum
_{\langle\langle i,j\rangle\rangle}Dv_{ij}\hat{z}\cdot\vec{S}_{i}\times\vec
{S}_{j} \label{eq1}%
\end{equation}
on a 2D hexagonal lattice of spins $\vec{S}_{i}$ with Heisenberg nearest
neighbor exchange interaction $J$ and Dzyaloshinskii-Moriya interaction $D$
(DMI) \cite{Dzyaloshinskii1958,Moriya1960}. The lattice sites $i$ and $j$ in
the second term of the right-hand side are next-nearest neighbors (NNN),
$v_{ij}=+(-)1$ on the upward (downward) pointing NNN triangle, as sketched in
Fig. \ref{fig1}(a). This Hamiltonian can support edge modes in the gap because
the nontrivial global phase or Berry curvature leads to bands with nonzero
Chern numbers ($\pm1$, depending on the sign of $D$) \cite{Xiao2010}. Each
spin quantum number $S$ is defined by $\vec{S}^{2}=\hbar^{2}S\left(
S+1\right)  .$ Raising and lowering operators read $S^{\pm}=S_{x}\pm iS_{y},$
respectively. According to the Holstein-Primakoff (HP) transformation
$S^{+}=\sqrt{2S}a^{\dag}\left(  1-a^{\dag}a/\left(  2S\right)  \right)
^{\frac{1}{2}}$, $S^{-}=\sqrt{2S}\left(  1-a^{\dag}a/\left(  2S\right)
\right)  ^{\frac{1}{2}}a$ and $S_{z}=S-a^{\dag}a,$ in terms of boson creation
($a^{\dag}$) and annihilation operators ($a$). When the number of magnons
$\sum_{i}a_{i}^{\dag}a_{i}$ is a small fraction of $NS$, where $N$ is the
number of spins, we can expand the square roots as $S^{+}\approx\sqrt
{2S}a^{\dag}$ and $S^{-}\approx\sqrt{2S}a$. However, the \textquotedblleft
participation\textquotedblright\ $\langle\Psi_{s}|a_{i}^{\dag}a_{i}|\Psi
_{s}\rangle$ of a localized soliton mode $|\Psi_{s}\rangle$ can become of the
order of $S$, even when the total magnon number is small. $\left(  1-a^{\dag
}a/\left(  2S\right)  \right)  ^{\frac{1}{2}}$ can then not be approximated by
unity, i.e., nonlinear terms in the Hamiltonian must be considered.

Other non-linearities can be added to the Haldane Hamiltonian Eq.~(\ref{eq1}),
such as an on-site magnetic anisotropy. Here we consider uniaxial
perpendicular magnetic anisotropy $\sum_{i}{KS_{z,i}^{2}}$. When $K$ is
negative (positive), the anisotropy is of the easy-axis (easy-plane) type. {{The Zeeman energy for a magnetic field $\vec{H}=\left(
0,0,H_{ext,z}\right)  $ is $H_{Z}=g\mu_{B}\sum_{i}{H_{ext,z}S_{z,i}}=g\mu
_{B}\sum_{i}{H_{ext,z}S-H_{ext,z}a_{i}^{\dag}a_{i}}$, where $g$ is the
g-factor, $\mu_{B}$ is the Bohr magneton. When the ground state magnetization
is aligned $\Vert$$\mathbf{\hat{z}}${, $H_{Z}$ rigidly shifts the entire spin
wave dispersion. This is always the case for perpendicular crystalline
anisotropy, while }${H}$$_{ext,z}$ should be sufficiently large when the
anisotropy is easy-plane. The spin-orbit interaction is an essential ingredient of our theory by generating a gap in the magnon spectrum. But in practice is rather weak. We find for realistic values of the parameter $D$ that the canting of spins relative to the quantization axis $\hat z$ is so small that it can be safely disregarded, e.g., in discussing the magnetic field dependence of the computed results.}} Magneto-dipolar
interactions are disregarded under the assumption that magnons with wave
lengths larger than the exchange length do not play a dominant role.

Consolidating the above: we work with the Hamiltonian
\begin{align}
H_{T} & =-JS\sum_{\langle i,j\rangle}(a_{i}^{\dag}a_{j}+H.c.)+  & \nonumber\\
& (3JS-2KS-g\mu
_{B}H_{ext,z})\sum_{i}a_{i}^{\dag}a_{i}  & \nonumber\\
& -DS\sum_{\langle\langle i,j\rangle\rangle}(iv_{ij}a_{i}^{\dag}a_{j}%
+H.c.)+H_{NL},  \label{eq2}%
\end{align}
with non-linearities to the fourth order in field operators :%
\begin{equation}
\begin{aligned}
\label{eq2_1}
H_{NL}  &  =\frac{J}{4}\sum_{\langle i,j\rangle}\left[  \left(  a_{i}%
n_{i}a_{j}^{\dag}+a_{i}n_{j}a_{j}^{\dag}+a_{i}^{\dag}n_{i}a_{j}+a_{i}^{\dag
}n_{j}a_{j}\right)\right. \\
& \left. -2a_{i}a_{j}^{\dag}-n_{i}n_{j}\right]-\frac{D}{4}\sum_{\langle\langle i,j\rangle\rangle}iv_{ij}\left[\left(a_{i}^{\dag
}n_{i}a_{j}+a_{i}^{\dag}n_{j}a_{j}\right)\right. \\
&\left. -\left(a_{i}n_{i}a_{j}^{\dag}+a_{i}n_{j}a_{j}^{\dag}\right)-2a_{i}a_{j}^{\dag}\right]  +\sum_{i}K(n_{i})^{2},
\end{aligned}
\end{equation}
where the occupation number $n_{i}=a_{i}^{\dag}a_{i}$. {The choice $-2KS=g\mu_{B}H_{ext,z}$ takes care of the
alignment of the spins as well as the band edge for all parameters which helps
to interpret the calculated phase diagrams.} (With the exception of
the term $n_{i}n_{j}$) the operators $n_{i(j)}$ appear in a \textquotedblleft
sandwiched\textquotedblright\ form such as $a_{i}n_{i}a_{j}^{\dag}$. { We refer to the
non-linearities generated by the anisotropy $K$ as \textquotedblleft self-Kerr
effect\textquotedblright, since it generates a frequency shift proportional to
the number operator (that should not be confused with the magneto-optical Kerr
effect).}

Rather than attempting to diagonalize this Hamiltonian, we iteratively search
for self-consistent soliton solutions. We start with a localized initial trial
wave function (WF) $|\Psi_{0}\rangle$ with density $\langle\Psi_{0}|n_{i}%
|\Psi_{0}\rangle=P_{0}>0$ for a site $i=0$ deep in the bulk and zero
otherwise. We keep $P_{0}$ constant during subsequent iterations by requiring
$\sum_{i}\langle\Psi_{m}|n_{i}|\Psi_{m}\rangle=\sum_{i}P_{i}^{\left(
m\right)  }=P_{0},$ where $m$ is the iteration step. This $H_{NL}$ can be
rewritten in terms of real space spinors $\psi_{i}=(a_{i},a_{i}^{\dag})^{T}$
to become matrices $\mathbf{H}_{T}^{\left(  m\right)  }$ with dimension
$2N_{x}\times N_{y}$, where $N_{x}$ ($N_{y}$) is the number of sites in the
$x$ ($y$) axis):
\begin{equation}
\mathbf{H}_{T}^{\left(  m\right)  }=\sum_{i,j}\psi_{j}^{\dag}\mathcal{H}%
(P_{i}^{\left(  m-1\right)  },P_{j}^{\left(  m-1\right)  })\psi_{i}.
\label{eq3}%
\end{equation}
$\mathcal{H}(P_{i}^{\left(  m-1\right)  })$ depends on $|\Psi_{m-1}\rangle$ by
the nonlinear (non-bilinear) terms in $H_{NL}$. $\mathbf{H}_{T}^{\left(
m\right)  }$ is diagonalized and $|\Psi_{m}\rangle$ is chosen to be the
eigenstate with the highest overlap $\left\vert \langle\Psi_{m}|\Psi
_{m-1}\rangle\right\vert .$ Self-consistency is reached when the overlap
approaches unity by a certain criterion for a solution with $\forall_{i}%
P_{i}<2S$. {Solitons are self-localized, dispersion-less
wavepackets that exist in the energy gaps of band structures. The Haldane
model has one finite band gap that splits the density of states but also a
finite spectral width with zero density of states at high and low energies
(semi-infinite gaps) (see Fig. 1(b)). Solitons can exist in the three gaps,
but they can be topologically relevant only in the internal band gap that can
support edge modes. We limit our search to solutions with frequencies in the
bandgap and thereby discard possible solutions outside the band
edges.} 

This iterative method sometimes fails to converge to a single
solution. We can overcome that problem by implementing an auxiliary temporal
periodicity as described in Appendix \ref{app1}.

\subsection{\label{sec1_b} Results}

Figure \ref{fig1}(b) shows the band structure of the linearized spin
Hamiltonian Eq. (\ref{eq1}) for an infinitely long quasi-1D ribbon that is 80
lattice sites wide with staggered {(zigzag)} free {(open)} boundary condition at the edges. It
hosts a topologically nontrivial band gap of $6\sqrt{3}DS$ (see Sec.
\ref{secq_c}) with chiral edge modes. {We assume $S=10$, $J=0.1$ and $D=0.01$ in all of the calculations in this paper, unless otherwise stated}. We carry out the bulk soliton search in
the finite gap of a lattice with $N_{x}=40$ and $N_{y}=80$ and free boundary
condition in $x$ and $y$ direction, with armchair and staggered edges,
respectively. We first focus on non-linearities caused by the anisotropy and
non-interacting magnons, i.e. when $\forall_{i}\langle\Psi_{s}|a_{i}^{\dag
}a_{i}|\Psi_{s}\rangle\ll S$, and discuss the magnon-magnon interaction below.
The results of the soliton search can be summarized by a phase diagram as a
function of intensity $P_{0}$ and anisotropy $K$. The energy of a converged
solution $E_{s}$ is indicated by the color code of the side bar in Fig.
\ref{fig1}(c). Mathematically, the assumption of non-interacting magnons
remains valid for other parameter regimes by scaling $S$ up with coefficient
$\mathcal{C}>1,$ while scaling down $J$ and $D$ by $1/\mathcal{C}$.

The threshold $P_{0}=P_{c,1},$ marked by green-dotted lines in Fig.
\ref{fig1}(c) is the intensity above which we find self-stabilized soliton
solutions. The solitons in the first region just above $P_{c,1}$ have energies
close to the band edges. Their amplitudes (or \textquotedblleft wave
functions\textquotedblright, WF, intensity) are spatially relatively extended,
as seen in the right panel of Fig. \ref{fig1}(c). We note the large
differences for positive and negative anisotropy: The energies of the solitons
for $K>0$ ($K<0$) are closer to the lower (upper) band edge for smaller
$P_{0}$ but move to the upper (lower) band edge with increasing $P_{0}$. For
$K<0$ ($K>0$), the nonlinearity tends to localize (delocalize) the WFs; a
soliton mode exists by compensation of nonlinearity and diffraction, so the
effective mass at the band edge must be positive (negative), which is the case
at the upper (lower) edge. $K<0$ ($K>0$) can be referred to as focusing
(defocussing) non-linearities, respectively. The focusing nonlinearity can
lead to solitons in both continuous and discrete media, while the defocussing
nonlinearity supports solitons only in lattices with gaps. \cite{Eiermann2004}%
. This difference becomes important when non-local magnon-magnon interactions
are taken into account (see below).

A second threshold $P_{0}=P_{c,2}$ as indicated by the red-dash line in Fig.
\ref{fig1}(c) marks a very different phase boundary; there is a sharp change
in energy and is also observed for the equivalent Floquet lattice, see the
discussion below and in Sec. \ref{sec2}.

Above a third threshold (not shown) the iterative solution scheme fails to
converge, but oscillates between two (or more) states. This means that the
eigenfunction of the (mean-field) nonlinear potential induced by one soliton
is a different soliton with different energy, indicating \textquotedblleft
breathing\textquotedblright. This situation is an artifact of the choice of
the initial condition. We implement a numerical method based on an auxiliary
time periodic potential as described in Appendix \ref{app1} in order to
converge unphysical breathing modes to steady state soliton solutions.

\begin{figure}[t]
\includegraphics[width=0.5\textwidth]{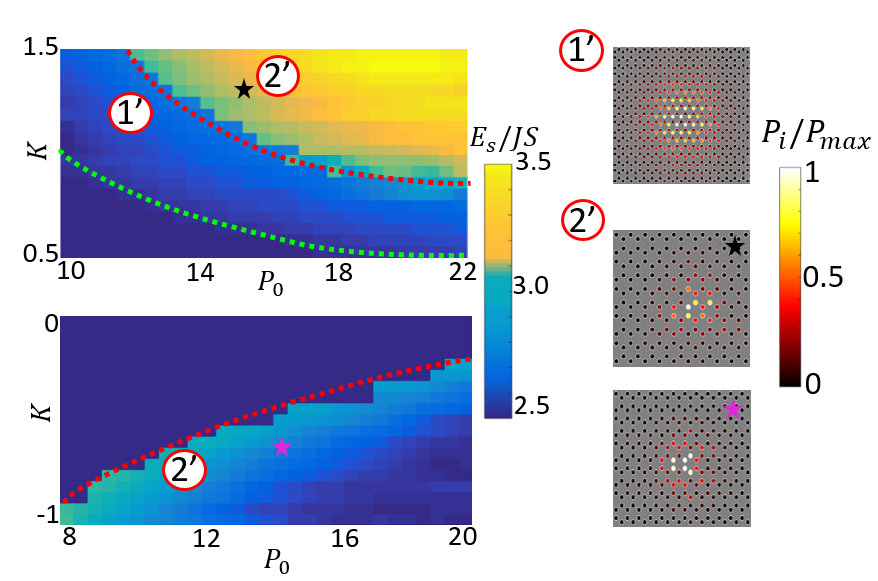}\caption{The bulk
soliton formation phase diagram for interacting magnons. The intensity
distribution of a soliton in region $1^{\prime}$ as well as two examples
(marked by black and purple stars) in region $2^{\prime}$ are shown. The green
(red) dashed line indicates $P_{c,1}^{\prime}$ ($P_{c,2}^{\prime}$). }%
\label{fig2}%
\end{figure}

The results in Figs. \ref{fig1}(c) hold when (for fixed $S/J$ and $K$) $S$ is
sufficiently large and the magnon interaction is small. In the following we
demonstrate how the higher order terms in the HP expansion proportional to $J$
and $D$ in Eq. (\ref{eq2_1}) modify the soliton phase diagram. The numerical
procedure is the same as before\textit{. }Figure \ref{fig2} illustrates a
first phase boundary at threshold $P_{c,1}^{\prime}$ for solitons as those in
region 1(2) of Fig. \ref{fig1}(c) for $K>0$ $\left(  K<0\right)  $,
respectively. For $K>0$, a phase boundary at $P_{c,2}^{\prime}$ (the red
dashed line in Fig. \ref{fig2}) similar to the one at $P_{c,2}$ in Fig.
\ref{fig1}(c) exists. Region $1^{\prime}$($2^{\prime}$) in Fig. \ref{fig2} is
similar to region $1$($2$). When (sufficiently strong) magnon-magnon
interactions are included, i.e. the terms proportional to $J$ and $D$ in Eq.
(\ref{eq2_1}), the phase boundary marking topological distinct phases exists
only when $K>0$ (see Fig. \ref{fig2}), because these interactions are of the
focusing type in continuous magnetic media \cite{Slavin1994,Stancil2009}. They
assist the local Kerr nonlinearity for $K<0$ but oppose it when $K>0$,
preventing formation of extended solitons in region $1^{\prime}$ to form for
$K<0$. Interaction also shifts the onset of soliton formation $P_{c,1}$ to
lower $P_{0}$ for the same $K<0$, while the topological change threshold moves
to higher $P_{0}$, i.e. $P_{c,2}^{\prime}>P_{c,2}$ for $K>0$.

{A scaling that extends the parameter space for which the present calculations are valid,
corresponds to keeping $S$, $K$, and $J/D$ constant, while scaling down $J$.
This decreases $P_{max}$ and therefore $P_{max}/S$. Approximating the envelope
cross sections of solitons of region 1 (1') by solutions of an effective 1D
nonlinear Schr\"{o}dinger equation \cite{Christodoulides1988,Kivshar1993}, we
deduce that $P_{max}$ is reduced by a factor 10 when $J/S$ is reduced by a
factor 100. Therefore, for $S=1$, $J=0.01$ and $D=0.001$, the phase diagrams
as Fig. \ref{fig1}(c) and Fig. \ref{fig2} hold for $\sim10$ times smaller $P_{0}$, $\sim100$
times smaller energies, while $P_{max}/S$ remains the same. We therefore can
extend our arguments to small spin systems when the exchange interaction is
small enough.}

\subsection{\label{secq_c} Discussion}

We now compare our results for local excitations in a wire to that of a
homogeneous excitation of the bulk system, following an analysis of the
Su-Schrieffer-Heeger (SSH) model \cite{Su1980}, i.e. spinless electrons on a
one-dimensional lattice with staggered hopping amplitudes, for which an
anharmonicity in the phonon amplitude causes a phase transition from a
topologically trivial to nontrivial phase \cite{Hadad2016}.

We can express the Haldane model in reciprocal space $H_{T}=\sum_{\vec{k}}%
\psi_{\vec{k}}^{\dag}\mathcal{H}_{k}\psi_{\vec{k}}$, where the spinor
$\psi_{\vec{k}}=(b_{\vec{k}},c_{\vec{k}})^{T}$ is written in terms of the
electron annihilation operators $b_{\vec{k}}$ and $c_{\vec{k}}$ on the two
sublattices. We can map the linearized magnetic Hamiltonian Eq. (\ref{eq2}) to
this form by
\begin{equation}
\mathcal{H}_{k}(\vec{k})=q_{0}I+\mathbf{q}(\vec{k})\cdot\vec{\sigma},
\end{equation}
where $q_{0}=3JS$, $I$ is the $2\times2$ unit matrix, $\vec{\sigma}%
=[\sigma_{x},\sigma_{y},\sigma_{z}]^{T}$ is the vector of Pauli matrices,
\begin{equation}
\mathbf{q}(\vec{k})=\left(
\begin{array}
[c]{c}%
-JS\sum_{i}\cos\left(  \vec{k}\cdot\vec{\lambda}_{\mathrm{NN},i}\right)  \\
-JS\sum_{i}\sin\left(  \vec{k}\cdot\vec{\lambda}_{\mathrm{NN},i}\right)  \\
-2DS\sum_{i}\sin\left(  \vec{k}\cdot\vec{\lambda}_{\mathrm{NNN},i}\right)
\end{array}
\right)  ,\label{eq6_1}
\end{equation}
and $\vec{\lambda}_{\mathrm{NN},i}$ ($\vec{\lambda}_{\mathrm{NNN},i}$) is one
of the three (i.e. $i=1,2,3$) lattice vectors to the NN (NNN) sites. The
energies $\mathcal{E}(\vec{k})$ and wave functions $|\Psi_{\vec{k}}\rangle$
for the two bands are
\begin{equation}
\mathcal{E}(\vec{k})=q_{0}\pm\sqrt{\mathbf{q}\cdot\mathbf{q}},\label{eq6}%
\end{equation}%
\begin{equation}
|\Psi_{\vec{k}}\rangle=\frac{1}{\sqrt{2}}\left(
\begin{array}
[c]{c}%
\sqrt{1\pm\frac{q_{z}(\vec{k})}{\sqrt{\mathbf{q}\cdot\mathbf{q}}}}\\
\mp\exp\left(  -i\tan^{-1}\frac{q_{y}(\vec{k})}{q_{x}(\vec{k})}\right)
\sqrt{1\mp\frac{q_{z}(\vec{k})}{\sqrt{\mathbf{q}\cdot\mathbf{q}}}}%
\end{array}
\right)  .\label{eq7}%
\end{equation}
The anisotropy is now represented by a $2\times2$ nonlinear Hamiltonian
$\mathcal{H}_{\mathrm{NL}}$ which is diagonal in the mean-field approximation
with components $[\mathcal{H}_{\mathrm{NL}}(\vec{k})]_{1,1}=KP_{u}\langle
\Psi_{\vec{k}}|b_{\vec{k}}^{\dag}b_{\vec{k}}|\Psi_{\vec{k}}\rangle$ and
$[\mathcal{H}_{\mathrm{NL}}(\vec{k})]_{2,2}=KP_{u}\langle\Psi_{\vec{k}%
}|c_{\vec{k}}^{\dag}c_{\vec{k}}|\Psi_{\vec{k}}\rangle$, where $P_{u}$ is the
density per unit cell of the homogeneously excited system. {Eqs. (\ref{eq6}),
(\ref{eq7}), and $\mathcal{H}_{\mathrm{NL}}$ define a self-consistent problem
that has to be solved numerically for each $\vec{k}$. At the Dirac points
$\vec{k}_{1}=[4\pi/3,0]$ and $\vec{k}_{2}=[2\pi/3,2\pi/\sqrt{3}]$ (a similar
discussion applies to the four other Dirac points), at which the gap opens and
closes, }$\mathbf{q}${$\left(  {\vec{k}_{1}}\right)  =3\sqrt{3}DS\mathbf{\hat
{z}}$ and }$\mathbf{q}$$\left(  {\vec{k}_{2}}\right)  =${$-3\sqrt
{3}DS\mathbf{\hat{z}}$. Therefore, $\mathcal{H}_{\mathrm{NL}}=KP_{u}%
(I\pm\sigma_{z})/2$, plus (minus) holding when only the upper (lower) bands
contribute, which is the case when $K$ is negative (positive). The
$\mathcal{H}_{\mathrm{NL}}$ therefore simply adds a mass coefficient
$\mathcal{M}=\pm KP_{u}/2$\textit{ }to $q_{z}$ in the unperturbed Hamiltonian.
It can be shown that the Chern number is $\pm1$ if $-6\sqrt{3}DS<KP_{u}%
<6\sqrt{3}DS$ and vanishes otherwise. Therefore, by increasing $P_{u}$ for any
sign of $K$, a transition from topologically nontrivial to trivial phase
occurs at a critical $P_{u,c}=6\sqrt{3}DS/|K|$. {Magnons
with focusing (defocusing) nonlinearity belong to a band with positive
(negative) effective mass, i.e. upper (lower) band, just as in insulators with
trivial bandgaps \cite{Cohen2003}. Therefore, the trivial mass term with
coefficient $\mathcal{M}$ as discussed above stems only from one of the bands,
depending on the sign of $K$.}

{When the mass in the core of the soliton changes from nontrivial to trivial,
the soliton mode should be exponentially localized in the area where the
trivial and nontrivial topology meet, i.e. at the edges
\cite{Hasan2010,Qi2011}. This is indeed evident from the spatial WF of
solitons of regions 1 (1') and 2 (2') (see Fig. \ref{fig1}(c) and Fig. \ref{fig2}): For the same
$P_{0}$, the energy of the more localized soliton is higher (lower) for
positive (negative) $K$, simply because the self-Kerr nonlinearity $\sum
_{i}K(n_{i})^{2}$ in Eq. (\ref{eq2_1}) indicates that $\left\vert K\right\vert \times
N_{s,1}\times(P_{0}/N_{s,1})^{2}>{\left\vert K\right\vert }\times
N_{s,2}\times(P_{0}/N_{s,2})^{2}$ and{ $N_{s,1}>N_{s,2}$, where $N_{s,1}$
($N_{s,2}$) is the number of sites with relatively large participation in the
soliton WF of region 1 or 1' (2 or 2'). The abrupt change of the boundary
conditions causes a jump in the formation energies by the non-trivial boundary
conditions.} In a homogeneously excited
system in which the Chern number is well defined, we find the transition at similar values of $P_{max}$ (see Fig. \ref{fig3_1}), which is another
indication that the soliton core is bulk-like.}

Figure \ref{fig3_1} shows the peak intensity $P_{max}=\max\langle\Psi_{s}%
|a_{i}^{\dag}a_{i}|\Psi_{s}\rangle$ as a function of $P_{0}$ for different
values of $K$ corresponding to the solitons of Fig. \ref{fig1}(c) and compares
it with the $P_{u,c}\left(  K\right)  $ for the bulk systems represented by
horizontal lines. When the integrated intensity of the soliton $P_{0}$ increases to
$P_{max}=P_{u,c}$ a phase change is expected. Indeed, at those powers
$P_{max}$ jumps to a higher level. In Fig. \ref{fig3_1}, $P_{max}<P_{u,c}$
when $P_{0}<P_{c,2},$ but $P_{max}>P_{u,c}$ for $P_{0}>P_{c,2}$. We therefore
conclude we can understand the phase boundary $P_{c,2}$ in terms of the
bulk-like mechanism in the soliton core. A similar relation for the bulk
systems can be obtained for the interacting system (not shown) with phase
change that occurs at $P_{c,2}^{\prime}$ of Fig. \ref{fig2}. Therefore, in
regions 1 and $1^{\prime}$, the maximum soliton WF intensity (and therefore
that at each site) is less than $P_{u,c}$, which means that the topology
remains non-trivial and the entire lattice has the same Chern number ($C=\pm
1$). In regions 2 and $2^{\prime}$, the soliton WF intensity at the center is
larger than $P_{u,c}$, hence the topology of the central part of the soliton
(trivial, $C=0$) and the rest of the lattice (nontrivial, $C=\pm1$) is different.

{If our interpretation is correct, edge modes around the
core of the solitons in region 2 (2') should generate a finite local density
of state (LDOS), which for a site ${i}${ reads \cite{Mesaros2010}}}
\begin{equation}
{{\rho_{i}(E)=\frac{-1}{\pi}\sum_{n}|\Psi_{n}^{\prime}(i)|^{2}}%
\mathrm{\operatorname{Im}}{\frac{1}{E-E_{n}+i\epsilon}}},
\end{equation}
{where the sum is over all self-consistent eigensolutions $\Psi_{n}^{\prime
}(i),E_{n}$ of $H_{T}$ (see Eq. (\ref{eq2})), and $\epsilon$ is a small broadening.
Figure \ref{fig4_new}(a) shows $\rho_{i}(E)$ for a soliton in region 1 (left) and region 2
(right) of Fig. \ref{fig1}(c), for $E$ inside the bandgap. The dominant peaks in
$\rho_{i}(E)$ agree with the soliton energies that are indicated by black
dashed lines. The soliton of region 2 is distinguished by additional two
relatively large peaks in the band gap that do not exist for the solitons {of
type 1. We define $\tilde{\rho}_{i}(E)=0$ for $E_{s}-\epsilon<E<E_{s}%
+\epsilon$ and {$\tilde{\rho}$}$_{i}(E)=\rho_{i}(E)$ otherwise, where
$\epsilon=JS/1000$; in other words we blend out t}he LDOS of the soliton in
order to enhance additional features. We then define $\rho_{M,i}$ as the
maximum value of {$\tilde{\rho}$}$_{i}(E)$ as a function of $E$ at each site
$i$. Figure \ref{fig4_new}(b) shows $\rho_{M,i}$ of the solitons in regions 1 and 2 as in
Fig. \ref{fig4_new}(a). Ignoring the sample boundaries, the excess LDOS is large in the
bulk (defining the soliton core) for the soliton phase 2 (but none is left in
1). Figure \ref{fig4_new}(b) is strong evidence that the soliton of region 2 (2') is
surrounded by edge modes, which supports the conjecture of a boundary between
two regions with different topology. In other words, the \textquotedblleft
mass transition\textquotedblright\ from nontrivial to trivial, i.e. the gap
opening by a trivial mass term $q_{z}=\mathcal{M}$ in the Hamiltonian is $\mathcal{M}\sigma_{z}$
or nontrivial one when $q_{z}$ is as in Eq. (\ref{eq6_1}),{ generates edge
states in the bulk around the soliton core of type 2 (2'). We note that
computational limitations force us to consider only small solitons, so Fig.
\ref{fig4_new}(b) does not resolve the ring-like LDOS expected for edge states.}}

We conclude that the soliton WF has a nonzero intensity in two regions with
different topology. Figures \ref{fig3_1_1}(a-b) illustrate this scenario.
Figure \ref{fig3_1_1}(a) is a sketch of the two possible situations, where the
non-trivial situation is the right panel and the interesting interface is
indicated by the red dotted line. {Figure 5(b) {shows the} spin wave
band structure for a topologically trivial and a nontrivial Haldane model with
the same gap, including the calculated Berry curvature $\vec{\Omega}\left(
\vec{k}\right)  ={\Omega}\left(  \vec{k}\right)  \hat{z}=\nabla_{\vec{k}%
}\times\langle\Psi(\vec{k})|i\nabla_{\vec{k}}|\Psi(\vec{k})\rangle$
\cite{Xiao2007,Xiao2010}\textbf{ }, while $C=\int_{\text{\textrm{BZ}}}%
dk^{2}\Omega(\vec{k})/$$\left(2\pi\right)  $, where BZ stands for the
first Brillouin zone. These band structures are global and cannot simply be
assigned to a small number of lattice sites such as the soliton core. Indeed,
assigning non-local properties such as a Chern number to a small number of
lattice sites such as the soliton core is not completely rigorous, but as our
discussion above indicates it is still a useful heuristic instrument.}

Appendix \ref{app2} analyzes the phase texture of the soliton WF's in support
of the above discussion.

\begin{figure}[b]
\includegraphics[width=0.5\textwidth]{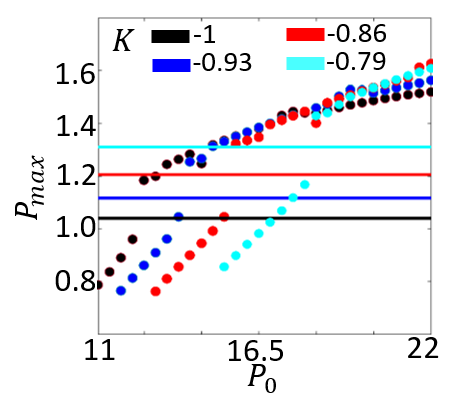}\caption{Peak intensity
$P_{max}$ for soliton formation calculated for four different values of the
anisotropy parameter $K$ as in Fig. \ref{fig1}(c). The horizontal lines
represent $P_{u,c},$ for the same $K$'s of homogeneously excited system, which
agree with the jumps in $P_{max},$ indicatig the distinct topology between
regions 1 and 2.}%
\label{fig3_1}%
\end{figure}

\begin{figure}[b]
\includegraphics[width=0.5\textwidth]{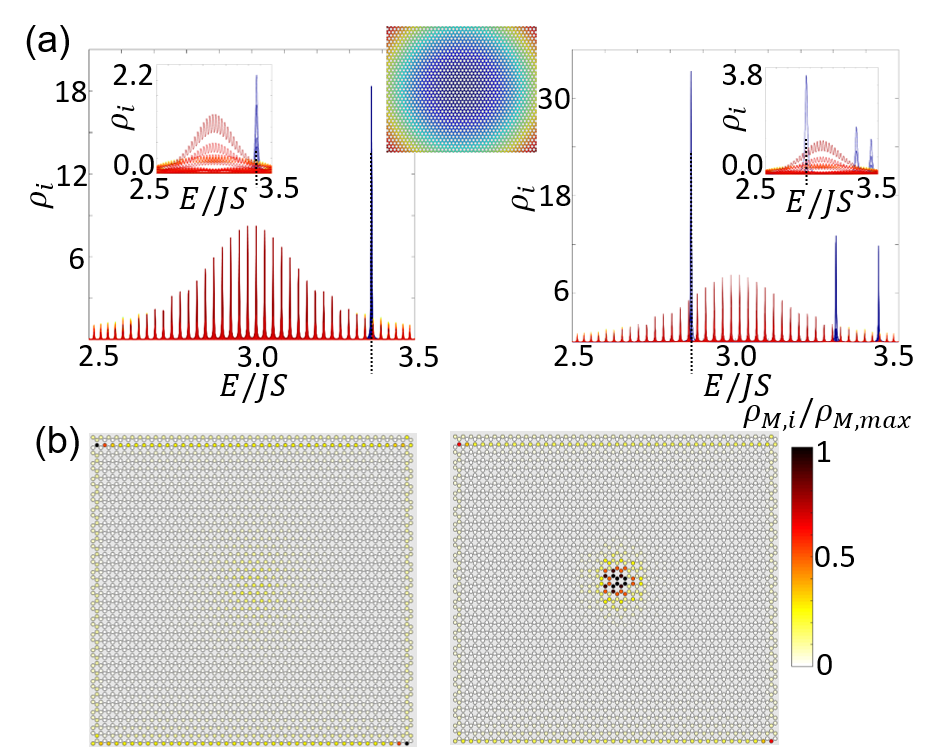}\caption{{(a) Density of
state ${\rho}_{i}(E)$ at each site $i$ for $E$ inside the band gap with
Lorentzian broadening $\epsilon=10^{-3}(10^{-2})JS$ in the main (inserted)
panels. The curves for all sites $i$ are plotted over each other colorcoded
according to the middle panel, i.e. line color is blue (red) for the sites in
the center (edges). { Left panel:
soliton in region 1 of Fig. \ref{fig1}(c) ($K=-1$ and $P_{0}=11.5$); Right
panel: soliton in region 2 of Fig. \ref{fig1}(c) ($K=-1$ and $P_{0}=12.9$). The black
dashed lines indicate the respective soliton energies. (b) $\rho_{M,i}=\max$%
}$\tilde{\rho}_{i}(E)$. Left panel: soliton in region 1 of Fig. \ref{fig1}(c) ($K=-1$
and $P_{0}=11.5$); Right panel: soliton in region 2 of Fig. \ref{fig1}(c) ($K=-1$ and
$P_{0}=12.9$).}}%
\label{fig4_new}%
\end{figure}

\begin{figure}[b]
\includegraphics[width=0.5\textwidth]{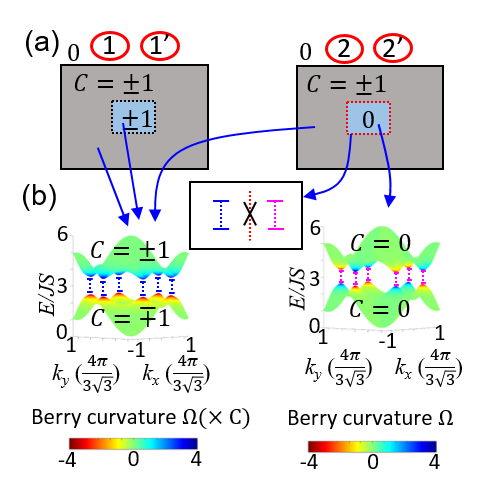}\caption{(a) The
topology of solitons on the Haldane lattice below and above $P_{u,c}$. The
blue region is the soliton core that at high intensities may have a different
Chern number $C$ from the rest of the lattice (embedded into a medium with
$C=0$). (b) {The bulk dispersion of the spin lattice for a
nontrivial ($D=0.01$ and $\mathcal{M}=0$) gap (left) and a trivial ($D=0$ and
$\mathcal{M}=3\sqrt{3}\times S/100$) gap (right). The color map encodes the
Berry curvature $\vec{\Omega}(\vec{k})=\Omega(\vec{k})\hat{z}$, for each band.
The arrows indicate the connection of different phases in (a) to the relevant
band structure panels in (b).}}%
\label{fig3_1_1}%
\end{figure}

\section{\label{sec2} Solitons in the Floquet magnonic Haldane model}

We now turn to the Floquet problem of the Haldane model with a periodic
potential in time. Solitons can exist in Floquet photonic topological
insulators, i.e. lattices of helical waveguides with on-site Kerr nonlinearity
($\sim n_{i}^{2}$ in Eq. (\ref{eq2_1})), in which the pitch of the helix is
the (spatial) Floquet period. Static bulk and propagating edge solitons have
been predicted\textbf{ }\cite{Lumer2013,Ablowitz2014,Leykam2016}. Here we
focus on \ the magnonic Haldane model with a time-periodic perturbation
\cite{Owerre2017}:
\begin{equation}
H_{F}=J\sum_{\langle i,j\rangle}S_{i}^{z}S_{j}^{z}+\frac{J}{2}\sum_{\langle
i,j\rangle}[S_{i}^{+}S_{j}^{-}e^{iA_{ij}\left(  t\right)  }+h.c.], \label{eq4}%
\end{equation}
where $A_{ij}$ is the Aharonov-Casher phase \cite{Aharonov1984} accumulated
upon hopping between nearest neighbors. It can be generated by elliptically
polarized light propagating normal to the lattice plane. The Hamiltonian in
Eq.~(\ref{eq4}) oscillates with period $T$ of the light field. The
non-topological Heisenberg Hamiltonian ($A_{ij}=0)$ can thus be driven to
generate a topologically non-trivial band structure. In other words, in
$\mathcal{T}\exp\left(  -i\int_{0}^{T}H_{F}\left(  t\right)  dt\right)
|\Psi\rangle=\exp\left(  -i\alpha T\right)  |\Psi\rangle$, where $\mathcal{T}$
is the time ordering operator and $|\Psi\rangle$ is the Floquet eigenstate,
the band structure that underlies the quasi-frequency $\alpha$ can be
topologically nontrivial, even when the band structure for the static part of
$H_{F}$ is trivial \cite{Kitagawa2010}. \textit{ }

We choose a periodic potential that gives rise to the Floquet equivalent of
the static Haldane model considered in Fig. 1 (see Appendix \ref{app3} for
details of the Floquet lattice \cite{Owerre2017}), which can be treated by the
soliton search method described in Appendix \ref{app1}\cite{Lumer2013}. Figure
\ref{fig4_1} shows the soliton quasi-energies $\alpha$ for $K=-1$ without
magnon-magnon interactions. Regions 1 and 2 of the static case (see Figs.
\ref{fig1}(c)) also exist for the Floquet problem (see Sec. \ref{secq_c}).

 {The similarity of the soliton phase diagrams for the static
and dynamically periodic system in the presence of the Kerr non-linearity
implies that including magnon-magnon interactions will modify the phase
diagram of the Floquet system in the same way as it affects the spatially
periodic one in Fig. \ref{fig2}.}

\begin{figure}[t]
\includegraphics[width=0.5\textwidth]{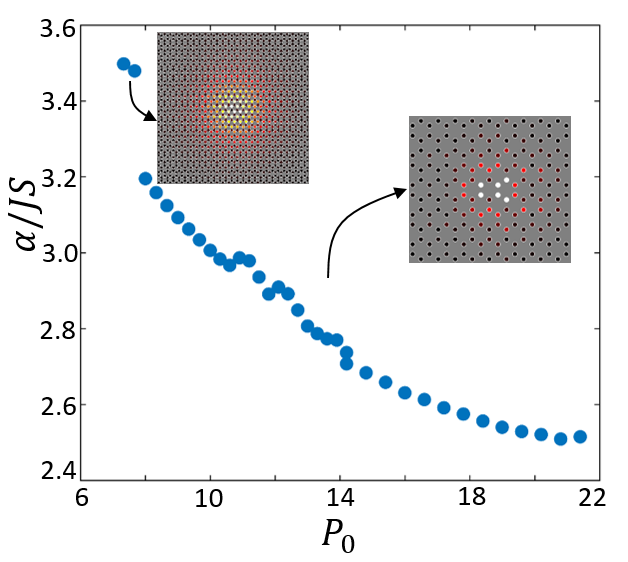}\caption{The quasi-energy
$\alpha$ (normalized by $JS)$ of a soliton in a Floquet lattice with $K=-1$ as
a function of $P_{0}$. Two exemplary intensity distributions also depicted.}%
\label{fig4_1}%
\end{figure}

\begin{figure}[t]
\includegraphics[width=0.5\textwidth]{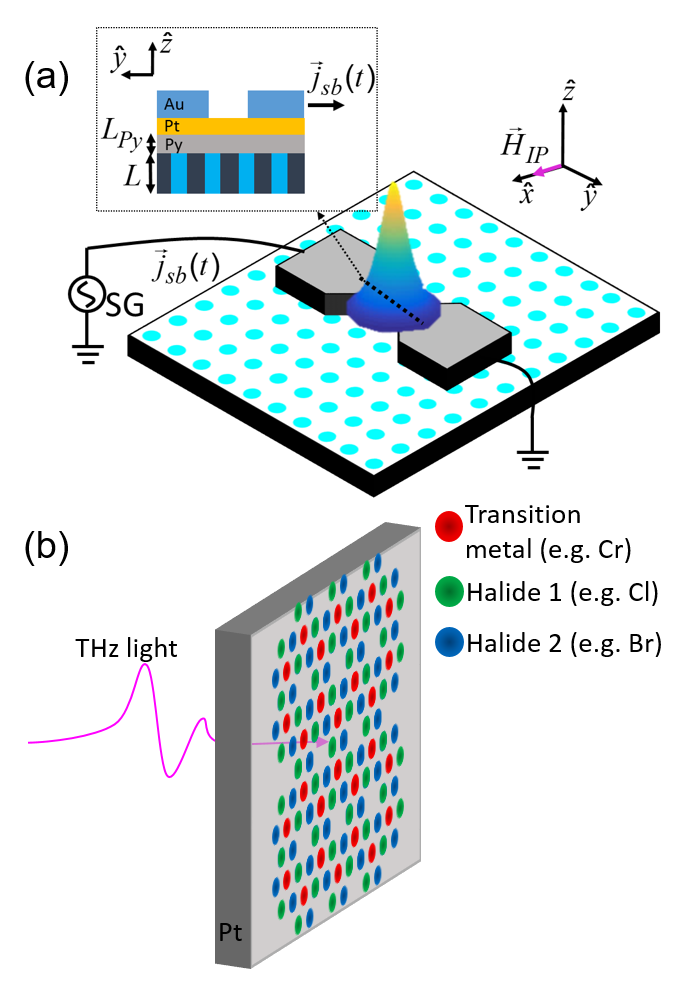}\caption{(a) The schematic
for an experimental realization of solitons described in this paper. SG stands
for signal generator. The current $\vec{j}_{sb}(t)$ passes through a current
line with a constricted region. A \textquotedblleft spin-wave bullet
(SWB)\textquotedblright\ wave function is shown schematically. The inset is a
$xz$ cross section, in which the materials of the SWB generation part (Au, Py,
and Pt), as well as thickness of Py ($L_{Py}$) and lattice ($L$) are pointed
out. The cyan color in both 3D and cross section panel indicate a magnetic
pillar array embedded in another magnetic host \cite{Shindou2013}. (b) Soliton
generation in a 2D transition metal halide proximate to a heavy metal (e.g.
Pt). A THz laser pulse irradiates the bilayer. The THz light excites magnons
in the 2D material.}%
\label{fig4_1b}%
\end{figure}

\section{\label{sec3}Experimental realization}

In the following we discuss the possibility of experimental realization of
magnonic solitons in topologically nontrivial lattices including their edges.
We realize that this is a larger order for the present stage of material
science, since the number of potential systems is limited and the required
structures might be difficult to fabricate. The real bottleneck could be the
dissipation and heating, which has been completely disregarded in the theory.
In order to keep these in check in excited systems, the materials and
structures must be grown and fabricated with high magnetic quality.

The magnonic solitons discussed here exist in a lattice that is strongly and
locally excited and has a topologically nontrivial band structure. Suggested
realizations are artificial magnonic crystals on the $\gtrsim$100 nm scale
\cite{Shindou2013}, but also natural materials such as the Kagome lattice of
$\text{Lu}_{2}\text{V}_{2}\text{O}_{7}$ \cite{Mook2015} or the hexagonal
lattice of $\text{Cr}\text{Br}_{3}$ \cite{Kim2016,Owerre2016}
\cite{Ideue2012,Jongh2001}. In magnetic films, solitons can be generated by
spin-Hall oscillators \cite{Liu2012,Demidov2012,Demidov2014} in which a point
contact of few tens of nanometer is deposited on top of a ferromagnet (e.g.
permalloy)/heavy metal (e.g. Pt) bilayer. The associated spin-orbit torques
cause self-limited spatially localized large-angle precessional states
referred to as \textquotedblleft spin-wave bullets\textquotedblright\ (SWB)
\cite{Demidov2012,Slavin2005,Kaka2005} with frequency and amplitude controlled
by the external magnetic field and charge current. In addition, precession can
be phase-locked to an ac modulated charge current drive \cite{Demidov2014}.

Figure \ref{fig4_1b}(a) sketches a device that could test our predictions. Two
triangular contacts of high-mobility metal film (e.g. from Au) force a focused
charge current $\vec{j}_{sb}(t)$ through the Pt, generating a SWB with
frequency $E_{s}/\left(  2\pi\hbar\right)  $ in the permalloy film by the spin
Hall effect. The latter is grown on a lattice of magnetic islands deposited in
the holes of another magnetic material \cite{Shindou2013}. The SWB generates
dynamic dipolar and exchange fields that excite the underlying lattice. The
existence and shape of the created lattice soliton can be studied using
spatiotemporal Brillouin-light scattering measurement technique
\cite{Demidov2008,Demokritov2008,Demidov2012,Bozhko2016} at
resolutions\textit{ }down to\textit{ }$\sim50$ nm
\cite{Demokritov2008,Demidov2012}. We give some estimates for this scenario in
Appendix \ref{app4}. We can operate the device also without the Py layer.
Current-induced self oscillations in the perpendicular magnetized material
then require an in-plane field \cite{Divinskiy2017}, but once excited, they
can be sustained by an ac current in Pt without the field.

Other interesting systems are 2D van der Waals materials such as
$\text{FePS}_{3}$, $\text{Cr}_{2}\text{Ge}_{2}\text{Te}_{6}$, and transition
metal trihalides, which have attracted attention for their tunable magnetic
properties \cite{Huang2017,Huang2018,Burch2018}. In the latter, the transition
metal (magnetic) sites form a hexagonal lattice, with a Heisenberg
(super)exchange interaction mediated by the halides. In addition, the magnetic
anisotropy can be tuned from in-plane to out-of plane by controlling the
halide composition: the magnetic anisotropy $K$ of $\text{CrCl}_{3-x}%
\text{Br}_{x}$ varies linearly with $x$ and changes from in-plane to out-of
plane at $x\approx2$ \cite{Abramchuk2018}. A DMI can be induced at the
interface to a heavy metal like Pt. The magnetic states can be studied by
X-ray magnetic circular dichroism (XMCD) spectroscopy with $\sim\mathrm{{nm}}$
($\sim50\,\mathrm{{fs}}$) spatial (temporal) resolution
\cite{Schutz1993,Kfir2015}. The magnon gap width of $\text{MX}_{3}$ is in the
range of $10\,\mathrm{{meV}}$ \cite{Klein2018}, i.e. we require THz excitation
for an efficient excitation of the spin system. Topological solitons can then
be generated in a monolayer of $\text{CrCl}_{3-x}\text{Br}_{x}$ on top of a
heavy metal irradiated by focused THz light with power above a certain
threshold, as sketched in Fig. \ref{fig4_1b}(b).

\section{\label{sec4}Conclusions}

In conclusion, we discuss the existence and characteristics of soliton
excitations in topologically nontrivial spin/magnonic lattices. We calculate
the soliton formation phase diagram in the presence of crystalline
anisotropies and magnon-magnon interactions. We understood in Sec.
\ref{secq_c} that without change in the topology of the underlying lattice,
topologically distinct solitons can form. We classify the phase diagrams and
predict a topological transition between soliton phases. Under certain
conditions we find a phase separation in the soliton itself. This implies the
existence of phase boundaries between solitons with trivial and non-trivial
topological properties. The boundary separates non-linear excited states with
trivial and non-trivial magnon gaps, which implies that at the boundary the
spin gap vanishes. Our toolbox to date does not allow to study the physical
consequences of such a topological interface for non-equilibrium systems, but
judging from the impact of the discovery of the interface states of electronic
topological insulators \cite{Hasan2010}, we expect interesting physics.

Such interplay of nonlinearity, dynamics, and topology has very recently
gained attention but is still widely unexplored
\cite{Hadad2016,Engelhardt2016,Peano2016}. Unfortunately, experimental
demonstration of a topologically nontrivial band gap of a magnonic lattice is
still lacking. Magnonic and other Bosonic edge states are not thermally
excited at low temperatures, which makes experimental observations
challenging. Non-linearities in highly excited systems can overcome this
problem \cite{Lumer2013,Ablowitz2014,Galilo2015,Leykam2016}. A soliton is a
non-dispersive mode and its formation is a threshold process. Finding our
predicted phase boundaries with sharp changes in soliton energies would be
strong evidence of a nontrivial topology of the underlying lattice. In
addition, the solitons phase texture as explained in Appendix \ref{app2}
reveals information about the topology of the underlying lattice and the class
of the soliton based on our classification. Solitons might serve for
information storage and transfer, and interaction/collision between solitons
as well as robust topological dynamic distinction, can be an information
resource. The possible unidirectional motion of the soliton when generated at
lattice edges \cite{Leykam2016} can lead to magnon squeezing
\cite{Kostylev2018} and enhanced optomagnonic coupling to photons
\cite{Osada2016,*Zhang2016,*Haigh2016,*Sharma2017}. We suggest that the
theoretical ideas can be realized and predictions tested in artificial
magnonic lattices as well as hybrid structures with single layer van der Waals ferromagnets.

\begin{acknowledgments}
M. Elyasi was supported by Postdoctoral Fellowship of Japan Society for the Promotion of Science (JSPS) for overseas researchers.
\end{acknowledgments}

\appendix

\section{\label{app1} Floquet calculation methods}

\begin{figure}[t]
\includegraphics[width=0.5\textwidth]{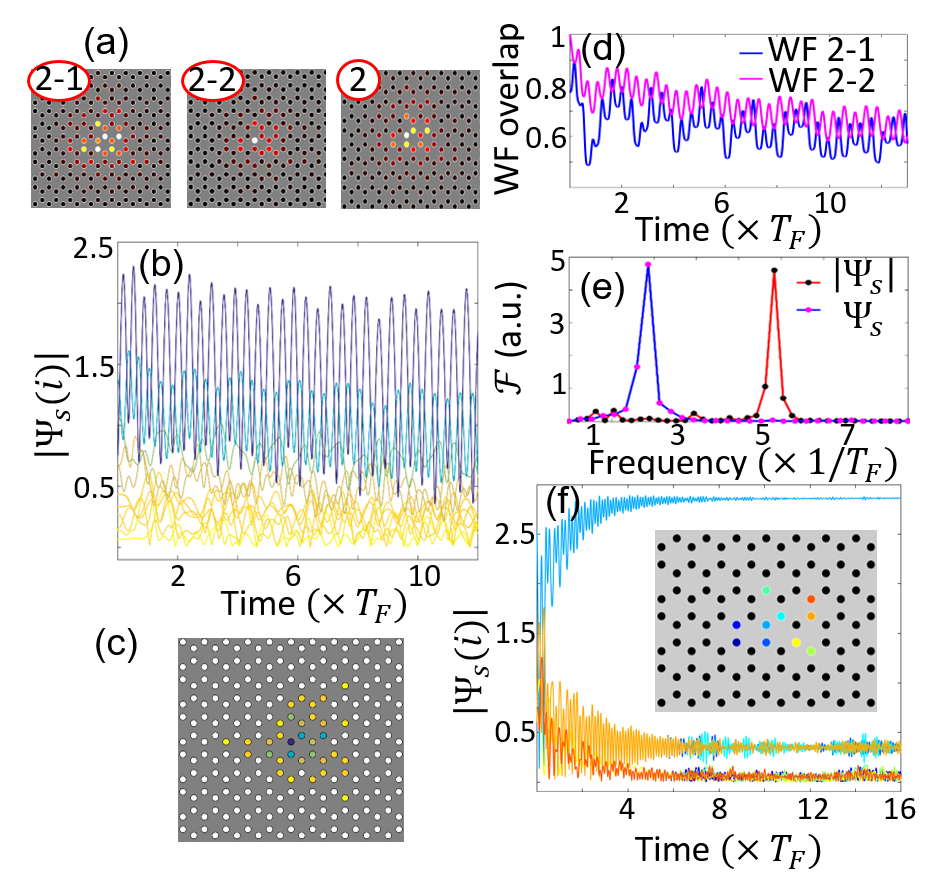}\caption{Soliton breathing
modes. (a) Two soliton WF's that are mean field solutions of each other's
potential, labeled 2-1 and 2-2. The\ converged stationary WF found with an
auxiliary Floquet potential as described in the text is labeled 2 (see also
Fig. \ref{fig1}(c)). (b) The time-dependent soliton WF $|\Psi_{s}(i)|$, for
sites $i$ with significant participation in the WF. The curves for the
presented sites are color coded in (c). The initial WF is the one
corresponding to the 2-2 in (a). The results are for $K=-1$ and $P_{0}=16$,
and $E_{2-1(2)}$ in $T_{F}=2\pi/|E_{2-1}-E_{2-2}|$ is energy of 2-1(2-2) in
(a). (c) The color code of the sites presented in (b). (d) The temporal
evolution of the soliton WF overlap with the WF's corresponding to 2-1 and
2-2. (e) The effective amplitude of the Fourier transform of the WF and its
amplitude. (f) Similar to (b) but for coarser time steps. The inset shows the
color coding of the sites.}%
\label{fig4}%
\end{figure}

As explained briefly in Sec. \ref{sec1_b}, in regions 2 and $2^{\prime}$ of
parameter space the iterative solution scheme sometimes fails to converge, but
oscillates between two (or more) states. This means that the eigenfunction of
the (mean-field) nonlinear potential induced by one soliton is another soliton
with different energy and vice versa, which indicates a breathing mode. Figure
\ref{fig4}(a) shows example WF's of such a breathing mode, labeled as 2-1 and
2-2. Adopting a time-periodic (Floquet) potential that oscillates between the
two solutions of 2-1 and 2-2 with natural period $T_{F}=2\pi/\left\vert
E_{2-1}-E_{2-2}\right\vert $ leads to a converged time-independent solution as
shown in Fig. \ref{fig4}(a) with label 2. In the following, we discuss the
rationale for $T_{F}$ and the method to solve the corresponding Floquet problem.

Figure \ref{fig4}(b) shows the temporal evolution of the initial WF labeled by
2-1, for several neighboring sites (with colors explained in Fig.
\ref{fig4}(c)). The WF amplitude $|\Psi_{s}|$ reflects the oscillation of the
nonlinear potential, which for the anisotropy term is $K|\Psi_{s}|^{2}$. The
normalized overlap of the 2-1 and 2-2 WFs in Fig. \ref{fig4}(d) reveals an
out-of phase oscillation with fixed frequencies that appears to be a mixture
of these two WF's. We test the assertion that $T_{F}=2\pi\hbar/\left\vert
E_{2-1}-E_{2-2}\right\vert $ by plotting $\mathcal{F}_{1\left(  2\right)
}=\sqrt{\sum_{i}|\mathcal{G}_{1\left(  2\right)  }(f)|^{2}}$ in Figure
\ref{fig4}(e), where the sum is over the sites of the lattice and $f$ is the
frequency. $\mathcal{G}_{1\left(  2\right)  }(f)$ is the Fourier transform of
$\Psi_{s}-\bar{M}_{1}$ $\left(  \left\vert \Psi_{s}\right\vert -\bar{M}%
_{2}\right)  $, where $\bar{M}_{1(2)}$ is the time average of $\Psi_{s}$
($|\Psi_{s}|$). The frequency of the main oscillatory features are above
$1/T_{F}$, supporting the assumption.

The time dependence for a coarser time steps is shown in Fig. \ref{fig4}(f) It
can be seen, that the WF amplitude converges to a fixed point with energy
below the bottom of the band; therefore the time steps chosen for evolution
should be fine enough for convergence to the solitons inside the finite gap.

We now posit that a periodic potential in time with periodicity $T_{F}$ can
bridge the \textquotedblleft breathing\textquotedblright\ and help finding a
converged static WF. $U(T_{F})=\mathcal{T}\exp\left(  -i\int_{0}^{T_{F}}%
H_{p}(t)dt\right)  $ is the evolution operator for time $T_{F}$ under a
periodic Hamiltonian $H_{p}(t)$ and $\mathcal{T}$ is the time ordering
operator. In the following we transform $U(T_{F})$ into $e^{-iH_{p,eff}T_{F}}%
$, where $H_{p,eff}$ is a static Hamiltonian. The band structure of the system
is then governed by $H_{p,eff}$ for integer multiples of $T_{F}$ with
eigenvalues or quasi-energies $\alpha$. This is then a static equivalent of
the Floquet $H_{p}$ with energy band structure $E=\alpha$.

The Floquet Hamiltonian can be written in terms of its discrete Fourier
components as $H_{p}(t)=\sum_{n=-\infty}^{\infty}H_{p}^{(n)}e^{in\omega_{pt}%
t}$, where $\omega_{pt}=2\pi/T_{F}$ and $H_{p}^{(n)}=\int_{0}^{T_{F}}%
H_{p}(t)e^{-in\omega_{pt}t}dt/T_{F}$. The effective static Hamiltonian,
$H_{p,eff}$, can then be written as a perturbation expansion
\cite{Scholz2010}
\begin{equation}
H_{p,eff}=H_{p,eff}^{(1)}+H_{p,eff}^{(2)}+H_{p,eff}^{(3)}+\dots,
\label{eq_a1_1}%
\end{equation}%
\begin{equation}
H_{p,eff}^{(1)}=H_{p}^{(0)},H_{p,eff}^{(2)}=\frac{1}{2}\sum_{n\neq0}%
\frac{[H_{p}^{(-n)},H_{p}^{(n)}]}{n\omega_{pt}}, \label{eq_a1_2}%
\end{equation}%
\begin{align}
H_{p,eff}^{(3)}  &  =\frac{1}{2}\sum_{n\neq0}\frac{[[H_{p}^{(n)},H_{p}%
^{(0)}],H_{p}^{(-n)}]}{n^{2}\omega_{pt}^{2}}\nonumber\\
&  +\frac{1}{3}\sum_{n^{\prime},n\neq0}\frac{[H_{p}^{(n)},[H_{p}^{(n^{\prime
})},H_{p}^{(-n-n^{\prime})}]]}{n^{\prime}n\omega_{pt}^{2}}. \label{eq_a1_3}%
\end{align}
Higher order terms $H_{p,eff}^{(n>3)}$ can be disregarded for large enough
$\omega_{pt}$, as is the case in our calculations.

In the soliton search, we start with either 2-1 or 2-2 Floquet WF's at $t=0$,
$|\Psi_{F}(t=0)\rangle$. Subsequently, we evolve $|\Psi_{F}(t=0)\rangle$ using
$H_{T}$ for one period, $T_{F}$, using the split-step method by updating the
non-linearities during the evolution. The resulting $|\Psi_{F}(0\leq t\leq
T_{F})\rangle$ is only a Floquet WF when $|\Psi_{F}(0)\rangle=|\Psi_{F}%
(T_{F})\rangle$. We diagonalize the resulting time-dependent Hamiltonian by
choosing an appropriate cut-off for $n$ in Eqs. (\ref{eq_a1_1}) to
(\ref{eq_a1_3}). The WF with the largest overlap with $|\Psi_{F}(0)\rangle$ is
chosen as the next $|\Psi_{F}(0)\rangle$ and the iteration goes on, until
$|\Psi_{F}(t)\rangle$ becomes WF of the nonlinearity-induced time-periodic
potential. In the calculations, we use $n$ up to 50.

Panel 2 in Fig. \ref{fig4}(a) is the converged result of the described method
for the initial breathing 2-1 and 2-2 WF's. The intensity distribution of 2 is
a mixture of 2-1 and 2-2, as expected. After convergence the distribution and
potentials are constant in time and the auxiliary periodic potential can be eliminated.

\section{\label{app2} Soliton phase texture}

\begin{figure}[b]
\includegraphics[width=0.5\textwidth]{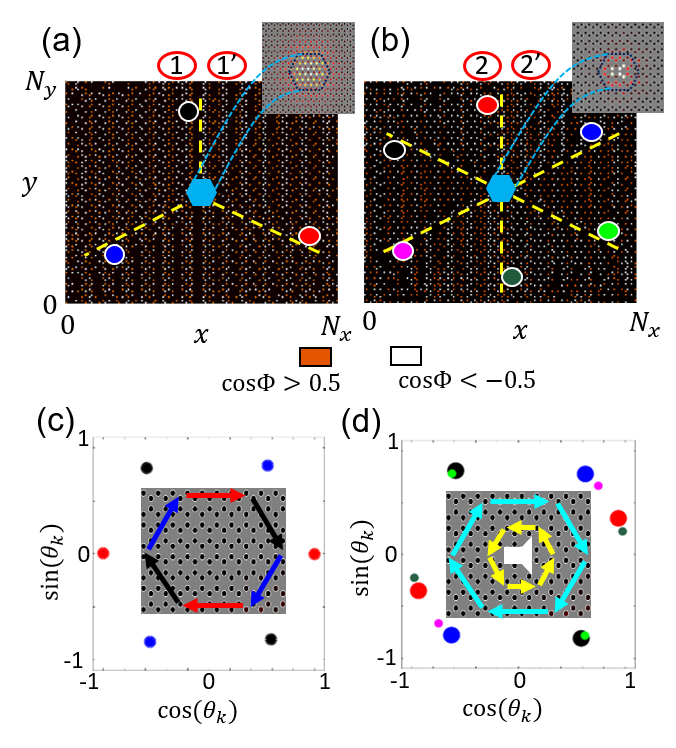}\caption{(a) and (b):
Spatially distributed $\cos\Phi$ of solitons centered at the origin (blue
hexagon) from regions1(1') and 2(2') in Figs. \ref{fig1}(c) and \ref{fig2}(a),
respectively. For better visibility only sites with $\cos\Phi>0.5$ and
$\cos\Phi<-0.5$ are shown. The insets depict the intensity distribution close
to the origin. The yellow dashed line separate the phase domains.The circles
label each domain with colors that are the same as that of the dots in (c) and
(d). (c) and (d): The dots indicate the calculated $\theta_{k}$ and
$\theta_{k}+\pi$ (Eq. (\ref{eqb1})) of each domain in (a) and (b),
respectively. The inner panels sketch the chirality of the gap modes in the
real-space lattice by arrows. In (c), these have the same color as the
corresponding $\theta_{k}$ and $\theta_{k}+\pi$. In (d), the chirality arrow
of the outer (inner) edge is depicted by cyan (yellow). The inner edge is
close to the origin of (b) where soliton intensity is maximized. Dots in (d)
have a different sizes for clarity and the \textquotedblleft
loudspeaker\textquotedblright\ mimicks the soliton amplitude snapshot in (b),}%
\label{fig_app2}%
\end{figure}We discuss now the soliton WF phase, $\Phi=\arg\Psi_{s}$ for the
topological transition in Figs. \ref{fig3_1} and \ref{fig3_1_1}(a-b). The
spatial dependence of the local precession phase, plotted in Figs.
\ref{fig_app2}(a-b) on a large scale around a soliton at the center, reveals a
global phase texture of either three or six domains. The \textquotedblleft
order parameter\textquotedblright\ is here a phase wave vector $\vec{k}_{\Phi
}$ defined as $\Phi\sim\operatorname{mod}[\vec{k}_{\Phi}\cdot\vec{r}+\phi
_{0},2\pi],$ where $\phi_{0}$ is a constant.

A soliton in a material with topologically trivial band gap, such as an
optical soliton with a vortex phase imprinted by the light shining on a
photonic lattice, can only have a scalar phase \cite{Neshev2004}. Its winding
number is a measure in real space and since the lattice is topologically
trivial, a non-zero value can only be imprinted by an external excitation.
However, when the winding (Chern) number of the Berry (geometric) phase in
momentum $\vec{k}$ space \cite{Xiao2010} is nonzero, the real space domains
are characterized by a vector $\vec{k}_{\Phi}$.

For a given domain, we can measure $\vec{k}_{\Phi}$ by the angle $\theta_{k}$%
\begin{equation}
\theta_{k}=\sum_{\vec{k}}\operatorname{mod}\left[  \tan^{-1}(\frac{k_{y}%
}{k_{x}}),\pi\right]  |\mathcal{G}_{\Phi}(\vec{k})|^{2}, \label{eqb1}%
\end{equation}
where $\mathcal{G}_{\Phi}\left(  \vec{k}\right)  $ is the normalized discrete
Fourier transform of the phase texture $\Phi\left(  \vec{r}\right)  =\arg
\Psi_{s}\left(  \vec{r}\right)  $. Here $\Psi_{s}$ is a stationary solution
with frequency $E_{s}/\left(  2\pi\hbar\right)  .$\ Both $\vec{k}_{\Phi}$ and
$-\vec{k}_{\Phi}$, i.e. $\theta_{k}$ and $\theta_{k}+\pi$ should be included
into the k-space summation, as demonstrated in Figs. \ref{fig_app2}(c-d), with
colored points corresponding to the domains in Figs. \ref{fig_app2}(a-b).

From Figs. \ref{fig3_1} and \ref{fig3_1_1}(a-b) and the corresponding
discussion follows that for regions 1 and $1^{\prime}$, the nontrivial
($C=\pm1$) and a trivial ($C=0$) topology meet at the edges to the vacuum,
whereas in regions 2 and $2^{\prime}$ a phase separation can exist within the
soliton as well. Therefore, the phase texture of the solitons of region 1 and
$1^{\prime}$ is solely determined by the outer edges, i.e. the minimal phase
texture should contain three domains each with mean Fourier components
deriving from two of the six Dirac points. The calculation of $\theta_{k}$ and
$\theta_{k}+\pi$ using Eq. (\ref{eqb1}) confirms this understanding as
depicted in Fig. \ref{fig_app2}(c). For the solitons of region 2 and
$2^{\prime}$, both the inner and outer edges contribute to the phase texture.
The inner edge is formed around the sites with $P_{i}>P_{u,c}$ (see Sec.
\ref{secq_c}), and its chirality is sketched by the yellow arrows in the inner
panel of Fig. \ref{fig_app2}(d) for the WF corresponding to Fig.
\ref{fig_app2}(b). We adopt arguments from Ref. [\onlinecite{Lumer2013}] to
determine the chirality of the inner edge. The phase texture of Fig.
\ref{fig_app2}(b) reflects these two counter propagating edges by six domains,
each with mean Fourier components occurring (approximately) at two of the six
Dirac points, as inferred from the values for $\theta_{k}$ and $\theta_{k}%
+\pi$ plotted in Fig. \ref{fig_app2}(d).

\section{\label{app3} Floquet lattice}

Eq. (\ref{eq4}) is the Hamiltonian of a 2D-lattice of Heisenberg
exchange-coupled local spins in the $xy$ plane when illuminated by circularly
polarized (CP) light with frequency $\omega$. We show here that this is
equivalent to a periodic static Haldane model.

A charge neutral particle with magnetic moment accumulates (Aharonov-Casher)
phase when moving with respect to an electric field \cite{Aharonov1984},
analogous to the Aharonov-Bohm phase for charged particles moving with respect
to magnetic fields \cite{Aharonov1959}. The accumulated phase upon hopping of
a magnon, $A_{ij}=g\mu_{B}\int_{\vec{r}_{i}}^{\vec{r}_{j}}\vec{A}(t)\cdot
d\vec{r}$ for $\vec{A}(t)=E_{0}(\pm\sin\omega t,\cos\omega t,0)/\omega$, is
$A_{ij}\propto g\mu_{B}E_{0}(\pm\cos(\omega t)\cos(\phi_{ij})+\sin(\omega
t)\sin(\phi_{ij}))/\omega$, where $E_{0}$ is the light electric field
amplitude, $g$ is the Land\'{e} $g$-factor, $\mu_{B}$ is Bohr magneton, and
$\phi_{ij}$ is the angle of the vector connecting site $i$ at $\vec{r}_{i}$ to
$j$ at $\vec{r}_{j}$.

Using Eqs. (\ref{eq_a1_1}) to (\ref{eq_a1_3}), the Hamiltonian Eq. (\ref{eq4})
can to leading order in the small parameter $\xi=g\mu_{B}E_{0}/\omega$ be
transformed into a time-independent sum of two contributions
\begin{equation}
H_{F,eff}^{(1)}=J\sum_{\langle i,j\rangle}[S_{i}^{z}S_{j}^{z}+\mathcal{J}%
_{0}(\xi)(S_{i}^{x}S_{j}^{x}+S_{i}^{y}S_{j}^{y})], \label{eq_app2}%
\end{equation}%
\begin{equation}
H_{F,eff}^{(2)}=\sum_{\langle\langle i,j\rangle\rangle}\sum_{n\neq0}%
\frac{(-1)^{n}J\mathcal{J}_{n}^{2}(\xi)}{n\omega}\sin(n\frac{2\pi}{3}%
v_{ij})\vec{S}_{k}\cdot\left(  \vec{S}_{i}\times\vec{S}_{j}\right)  ,
\label{eq_app3}%
\end{equation}
where $k$ is a site between two next nearest neighbors (NNN) $i$ and $j$, and
$\mathcal{J}_{n}$ is the $n$'th order Bessel function of the first kind.
Focussing on a perpendicular equilibrium magnetization, $\vec{S}_{k}$ can be
written as a sum of a static and dynamic contribution as $\vec{S}_{k}%
=S_{0,k}\hat{z}+\delta\vec{S}_{k}$ in Eq. (\ref{eq_app3}), which to leading
order reduces to a term similar to the DMI\ in Eq. (\ref{eq1}) with effective
DMI coefficient $D_{F}=-\sqrt{3}JS\mathcal{J}_{1}^{2}(\xi)/\omega$,
\begin{equation}
H_{p,eff}=H_{F,eff}^{(1)}+\sum_{\langle\langle i,j\rangle\rangle}D_{F}%
v_{ij}\hat{z}\cdot\left(  \vec{S}_{i}\times\vec{S}_{j}\right)
\end{equation}
where $\xi$ and $\omega$ are tunable by the power and frequency of the light.
We thereby recover the static Haldane model in the Floquet manner without
intrinsically broken inversion symmetry (and therefore DMI). Non-linearities
affect $H_{p,eff}$ in the same way as in the main text. Hence, the soliton
search procedure as explained in Appendix \ref{app1} can be applied, whereby
the Floquet period $T_{F}$ in Appendix \ref{app1} is fixed by $T=2\pi/\omega$.
Moreover, the initial trial WF of the iterative method is $\langle\Psi
_{F}(t=0)|n_{i}|\Psi_{F}(t=0)\rangle=P_{0}$ for a site $i=0$ in the bulk and
zero otherwise.

\section{\label{app4} Parameter estimates}

Magnons in crystals with periodic magnetization $M_{s}$ on length scales of
$d\sim0.5\,\mathrm{\mu m}$ are dominated by dipolar interactions. In a
structure consisting of a regular lattice of holes in a host magnet film that
are filled with sufficiently different magnetic materials and a filling
fraction $F\sim10^{-2},$ the Chern number of a specific band can be tuned
between zero, 1 and 2, by changing $d$ and the aspect ratio of the unit cell
\cite{Shindou2013}. The total spin in each unit cell is then $S=(M_{s,f}%
F+M_{s,l})d^{2}L/\left(  \gamma\hbar\right)  \approx M_{s,l}d^{2}L/\left(
\gamma\hbar\right)  $, where $-\gamma$ is the gyromagnetic ratio and subscript
$l$ ($f$) refers to the magnetic material of the lattice host (filling). $S$
can be tuned by the film thickness $L$, while the topological invariant is
kept constant, as long as the translational invariance along $\hat{z}$ is a
good assumption (thick film limit). We assume that we can locally excite the
lowest magnonic bands with constant amplitude along $\hat{z}$ by the dynamic
spin transfer torque of a spin wave bullet in a Py overlayer with in-plane
magnetization, as in Figure \ref{fig4_1b}(a).

Based on the typical frequencies of the band widths\textit{ }of dipolar
magnonic lattices (corresponding to $JS$ in the Haldane model) $\sim
10\,\text{GHz}$, for $L\sim10\,\mathrm{\mu}\text{m}$, i.e. $S\sim10^{15}$, we
have a correspondence to $J\sim0.0001$ in the Haldane model. The gap width and
topology of the magnonic lattice is governed by the aspect ratio of the
rectangular 2D unit cell with fixed area. It can be tuned to correspond to the
$D/J$ parameter used here. Therefore, the soliton phase diagram for $K=-1$ of
Fig. \ref{fig1}(c) can be achieved in a magnonic lattice with $K\sim-0.001$ by
a crystalline anisotropy constant $K_{u}$ through $K=-2\gamma K_{u}/\left(
\mu_{0}M_{s,l}S\right)  $, or $K_{u}\sim10^{4}\,$\textrm{J/m}$^{3}$ for
$\gamma=2.2\times10^{5}\,\mathrm{A/(m\mathrm{\cdot}s)}$ and $M_{s,l}%
=10^{5}\,A/m$. For comparison, $K_{u}$ for YIG and $\text{L1}_{0}$ FePt, are
of the order of $\sim10^{3}$ and $\sim10^{6}\,\text{J}/\text{m}^{3}$,
respectively \cite{Bryant1988,Thiele1998}. Tuning $S$ by changing $L$ while
keeping all other parameters intact, thus should allow resolving soliton phase
diagrams with (Fig. \ref{fig2}) and without (Fig. \ref{fig1}(c)) magnon-magnon
interactions. A possible concrete choice would be YIG for the film, while the
filling material is Fe (or Co). An ability to change the crystalline
anisotropy from say $10^{3}$ to $10^{4}\,\text{J}/\text{m}^{3}$ by using
different crystal growth directions, doping, or gating of YIG, could be useful
to map the soliton phase diagrams.

The spin-Hall oscillators that generate self-localized oscillations (SWB) can
be fabricated with spatial half-widths at half-maximum $R_{SWB}$ of
$\sim500\,\text{nm}$ \cite{Demidov2012}, which would cover approximately a
unit cell of the lattice in the previous example, which is sufficient to
generate the localized solitons predicted in Figs. \ref{fig1}(c) and
\ref{fig2}. In a $\sim10\,\mathrm{\mu}$\textrm{m} wide disk Py/Pt bilayer the
SWB were excited for an in-plane equilibrium magnetization. For a fixed
current above threshold, an in-plane magnetic field $\vec{H}_{IP}$ of
$4\times10^{4}-12\times10^{4}\,\text{A/m}$ can tune the SWB frequencies in the
range of $5-10\,\text{GHz}$. The perpendicular magnetization of the underlying
magnonic lattice is not significantly affected when the perpendicular
anisotropy is strong enough, e.g. $K_{u}\sim10^{4}\,\mathrm{J/m}^{3}$ while
the in-plane magnetization of Py is stabilized by the thin film shape anisotropy.

Modeling the SWB by a macrospin precession within the bullet volume, we can
calculate the effective dipolar field on the lattice below the SWB as
$h_{SWB,l}>\pi R_{SWB}^{2}L_{\mathrm{Py}}M_{s,\mathrm{Py}}\sin\theta
_{c}/(r_{w}^{2}L)$, where $L_{\mathrm{Py}}$ is the Py layer thickness
($5\,$nm), $M_{s,\mathrm{Py}}=7\times10^{5}\,\text{A/m.}$ $\theta_{c}$ is the
precession cone angle, and $2r_{w}$ is the diameter of the expected lattice
soliton. With $r_{w}\approx0.1\,\mathrm{\mu m}$, $\theta_{c}=\pi/4$,
$h_{SWB,l}>2.5\times10^{3}\,\text{A/m}$. A magnon mode in the lattice resonant
with the frequency $\omega_{SWB}$ of the driving field $h_{SWB,l}$ is
estimated as $n_{l}\sim(2h_{SWB,l}\gamma^{2}R_{SWB}^{2}L\sqrt{M_{s,l}%
/2\gamma\hbar}/\zeta_{m})^{2}$, where $\zeta_{m}$ is the magnetic damping,
which in a best case scenario $\sim1\,\text{MHz}$ for YIG. Therefore,
$n_{l}\sim1.5\times10^{15}$, which is $\sim1-10\,S$ that is in the range of
$P_{0}/S$ of Figs. \ref{fig1}(c) and \ref{fig2}. By adopting $M_{s}$ of the
host material YIG (which is smaller than the filling materials Fe or Co), this
is a lower bound for $n_{l}$. It should be noted that $h_{SWB,l}$ and
consequently $n_{l}$ can be tuned by the charge current amplitude and the
locking microwave charge current power \cite{Demidov2014}. In the steady state
the charge current generates an SWB that stabilizes the solitons by
regenerating the losses due to damping. $\tilde{E}$

\end{document}